\pdfoutput=1
\documentclass[aps,pre,reprint,superscriptaddress]{revtex4-2}
\usepackage[colorlinks=true,allcolors=blue]{hyperref}
\usepackage{bm}
\usepackage{newtxtext}
\usepackage{newtxmath}
\usepackage{anyfontsize}
\usepackage{amsmath}

\begin{document}
\title{Effective Langevin equations leading to large deviation function of time-averaged velocity for a nonequilibrium Rayleigh piston}
\author{Masato Itami}
\email{itami@r.phys.nagoya-u.ac.jp}
\affiliation{Department of Physics, Nagoya University, Nagoya 464-8602, Japan}
\author{Yohei Nakayama}
\email{r\_nakayama@tohoku.ac.jp}
\affiliation {Department of Applied Physics, Tohoku University, Sendai 980-8579, Japan}
\author{Naoko Nakagawa}
\email{naoko.nakagawa.phys@vc.ibaraki.ac.jp}
\affiliation {Department of Physics, Ibaraki University, Mito 310-8512, Japan}
\author{Shin-ichi Sasa}
\email{sasa@scphys.kyoto-u.ac.jp}
\affiliation{Department of Physics, Kyoto University, Kyoto 606-8502, Japan}
\date{\today}

\begin{abstract}
We study fluctuating dynamics of a freely movable piston that separates an infinite cylinder into two regions filled with ideal gas particles at the same pressure but different temperatures.
To investigate statistical properties of the time-averaged velocity of the piston in the long-time limit, we perturbatively calculate the large deviation function of the time-averaged velocity.
Then, we derive an infinite number of effective Langevin equations yielding the same large deviation function as in the original model.
Finally, we provide two possibilities for uniquely determining the form of the effective model.
\end{abstract}

\maketitle

\section{Introduction}\label{sec:intro}

Large deviation functions have played a prominent role in statistical physics.
In equilibrium systems, large deviation functions of thermodynamic variables are given by the corresponding thermodynamic functions in the thermodynamic limit, which fully characterize thermodynamic properties of the equilibrium systems~\cite{Ruelle1969,Lanford1973,Ellis1985,Touchette2009}.
For example, the large deviation function of the energy density for an isolated system is equivalent to the thermodynamic entropy density, which is known as Einstein's fluctuation theory.
To investigate and characterize nonequilibrium systems, large deviation functions of time-averaged currents have been studied~\cite{Oono1989,Bertini2007,Derrida2007,Touchette2009}.
One of the notable results of the last two decades is that a symmetry property of the large deviation function of time-averaged entropy production rate, generally valid far from equilibrium, was derived as a result of the time-reversal symmetry of microscopic mechanics, which is called the fluctuation theorem~\cite{Evans1993,Gallavotti1995,Jarzynski1997,Kurchan1998,Lebowitz1999,Maes1999,Crooks1999,*Crooks2000,Jarzynski2000,Seifert2005,*[{For a review, see }][{}] Seifert2012}.
This symmetry property enables us to easily derive the Green--Kubo relations, the McLennan ensembles, and the Kawasaki nonlinear response relation~\cite{Crooks2000,Hayashi2006}.
The symmetry property is also useful for estimating forces acting on motor proteins~\cite{Hayashi2018,Hasegawa2019}.

From a theoretical viewpoint, the large deviation functions of time averaged quantities cannot be easily calculated in microscopic many-body systems.
Here, there have been cases that an effective stochastic description at some coarse-grained level precisely provides the large deviation function~\cite{*[{For a review, see }][{}] Bertini2015}.
One may analyze such stochastic models to argue some properties of the large deviation functions without considering the connection to the microscopic description.
This approach is useful, in particular, when seeking universal properties independent of microscopic details. 
However, when we are interested in microscopic mechanisms for properties described by the large deviation functions, the connection to the microscopic description also should be understood.
The main purpose of this paper is to clarify the relationship among the microscopic description, the effective mesoscopic description, and the large deviation function for the simplest example, a nonequilibrium version of the Rayleigh piston model~\cite{vanKampen1961,Alkemade1963,vanKampen2007,Sarracino2013}.
This model is mainly used to investigate the adiabatic piston problem~\cite{Feynman1963,Callen1985,Lieb1999,Gruber1999I,Gruber1999II,Gruber2006,Kestemont2000,Munakata2001,Chernov2002,Plyukhin2004,Meurs2004,MalekMansour2006,Cencini2007,Fruleux2012,Itami2015I,Caprini2017,Khalil2019}.

It has been known that effective stochastic models are formally derived by using the projection operator method~\cite{Zwanzig1961,Mori1965,Kawasaki1973} and the nonequilibrium statistical operator method~\cite{Bashkirov1970} from the microscopic description.
It is not obvious to check the assumption made in the derivation. 
For example, although the noise properties in the obtained equations basically depend on the choice of slow variables and the projection method, they were often physically assumed without studying the nature of slow variables. 
Still, for equilibrium systems, following the Onsager theory~\cite{Onsager1931I,*Onsager1931II,Onsager1953}, we can restrict the form of the effective model by considering the regression hypothesis and the detailed balance condition.
However, this cannot be used for nonequilibrium systems due to the breakdown of the detailed balance condition.
Even for the simplest example studied in this paper, we do not find a consistent derivation method of the effective model from the microscopic description.

Now, putting aside the derivation of the effective stochastic model from a microscopic model, we study conditions for the effective model.
As a necessary condition, it should precisely reproduce the large deviation function of a time-averaged velocity of a piston for the microscopic model.
Such an effective model was discovered numerically in~\cite{Seya2020}.
Fortunately, the microscopic model is so simple that we can calculate the large deviation function in a perturbative expansion in powers of a small parameter $\epsilon$, where $\epsilon$ is the square root of the mass ratio of a light gas particle to a heavy piston.
We can also calculate the large deviation functions for effective models whose forms are assumed with undetermined parameters.
By comparing these two results, we find that an infinite number of effective models yield the same large deviation function as in the microscopic model to first order in $\epsilon$.
We then consider two possibilities for determining the effective model uniquely.

The remainder of this paper is organized as follows.
In Sec.~\ref{sec:model}, we introduce our setup and derive the dimensionless form of the equation we study.
In Sec.~\ref{sec:Pss}, we review basic properties of the model, which are already known from the previous works.
Next, we calculate the large deviation function of the time-averaged velocity in Sec.~\ref{sec:LD}.
In Sec.~\ref{sec:ELE}, we identify effective Langevin equations that reproduce the same large deviation function as in the microscopic model, and propose the conditions that uniquely determine the form of the Langevin equation.
The final section is devoted to a brief summary and some concluding remarks.

\section{Model}\label{sec:model}
We study the dynamics of a rigid piston of mass $M$ that separates an infinite cylinder into two regions filled with ideal gas particles of mass $m$.
The piston is freely movable in one direction inside the cylinder of cross-sectional area $S$.
Let $V$ denote the velocity of the piston.
The gas in the left and right regions are initially prepared at equal pressure $p$ but different temperatures $T_{\mathrm{L}}$ and $T_{\mathrm{R}}$, respectively.
Suppose, without loss of generality, that $T_{\mathrm{L}} < T_{\mathrm{R}}$.
The particles collide elastically and instantaneously with the piston only once, and then the particles are in equilibrium before colliding with the piston.
We model the collisions between the piston and particles by random events at a collision rate
\begin{align}
\lambda(v,V) &= \frac{pS}{k_{\mathrm{B}}T_{\mathrm{L}}}(v-V)\theta(v-V) f^{\mathrm{L}}_{\mathrm{eq}}(v) \notag\\
& \quad +\frac{pS}{k_{\mathrm{B}}T_{\mathrm{R}}}(V-v)\theta(V-v) f^{\mathrm{R}}_{\mathrm{eq}}(v)
\end{align}
with the Maxwell distribution
\begin{align}
f^{\mathrm{L/R}}_{\mathrm{eq}}(v)=\sqrt{\frac{m}{2\pi k_{\mathrm{B}}T_{\mathrm{L/R}}}}\exp\left(\frac{-mv^{2}}{2k_{\mathrm{B}}T_{\mathrm{L/R}}}\right),
\end{align}
where $v$ is the velocity of a colliding particle, $k_{\mathrm{B}}$ is the Boltzmann constant, and $\theta(\cdot)$ is the Heaviside step function.
Note that from the ideal gas law, $p/(k_{\mathrm{B}}T_{\mathrm{L/R}})$ equals the number density of the gas in the left/right region.
Using the laws of the conservation of energy and momentum, the transition probability density per unit time from $V$ to $V'$ is given by
\begin{align}
W(V'\vert V)=\lambda(v,V)\frac{\mathrm{d}v}{\mathrm{d}V'}
\end{align}
with
\begin{align}
v=\frac{M+m}{2m}V'-\frac{M-m}{2m}V.
\end{align}
Then, the time evolution of the probability density of $V$ at time $t$, $P(V,t)$, is governed by the following master-Boltzmann equation:
\begin{align}
\frac{\partial P(V,t)}{\partial t} &=\int\mathrm{d}V'\; \Big[ W(V\vert V')P(V',t)-W(V'\vert V)P(V,t) \Big] .
\label{eq:master}
\end{align}
This equation has been used to investigate the adiabatic piston problem~\cite{Gruber1999I,Gruber1999II,Gruber2006,Munakata2001,Meurs2004,Fruleux2012,Itami2015I}.
Using formally the Kramers--Moyal expansion~\cite{Risken1996,Gardiner2009}, we have
\begin{align}
\frac{\partial P(V,t)}{\partial t} = \sum_{n=1}^{\infty}\frac{(-1)^{n}}{n!}\frac{\partial^{n}}{\partial V^{n}}\Big[ a_{n}(V) P(V,t)\Big]
\label{eq:masterKM}
\end{align}
with
\begin{align}
a_{n}(V) &= \int\mathrm{d}V'\; (V'-V)^{n} W(V'\vert V)
\notag\\
&= \int\mathrm{d}v\; \left[ \frac{2m(v-V)}{M+m}\right]^{n}\lambda(v,V).
\label{eq:cKM}
\end{align}

Let us introduce a dimensionless small parameter
\begin{align}
\epsilon \equiv \sqrt{\frac{m}{M}} \ll 1,
\end{align}
which implies that the relaxation time of $V$ is much larger than the average time between collisions.
For later use, we introduce a dimensionless parameter defined by
\begin{align}
\phi\equiv\left(\frac{T_{\mathrm{R}}}{T_{\mathrm{L}}}\right)^{{1}/{4}}>1.
\end{align}
Let $T$ denote the geometric-mean temperature explicitly written as
\begin{align}
T\equiv\sqrt{T_{\mathrm{L}}T_{\mathrm{R}}}.
\end{align}
It will be shown later that $T$ is the kinetic temperature of the piston to lowest order in $\epsilon$.

We now introduce the rescaled dimensionless variables used in~\cite{Seya2020}
\begin{align}
\tau\equiv\frac{\epsilon pS(\phi+\phi^{-1})\,t}{\sqrt{\pi Mk_{\mathrm{B}}T/8}},\quad \mathcal{V}\equiv\frac{V}{\sqrt{k_{\mathrm{B}}T/M}}.
\label{eq:res_vari}
\end{align}
The probability density of $\mathcal{V}$ at $\tau$ is given by 
\begin{align}
\mathcal{P}(\mathcal{V},\tau)=P(V,t)\frac{\mathrm{d}V}{\mathrm{d}\mathcal{V}}.
\end{align}
Using these new variables, we rewrite (\ref{eq:masterKM}) with (\ref{eq:cKM}) as
\begin{align}
\frac{\partial \mathcal{P}(\mathcal{V},\tau)}{\partial\tau} = \sum_{n=1}^{\infty}\frac{(-1)^{n}}{n!}\frac{\partial^{n}}{\partial \mathcal{V}^{n}}\Big[ \alpha_{n}(\mathcal{V}) \mathcal{P}(\mathcal{V},\tau)\Big]
\label{eq:masterKMres}
\end{align}
with
\begin{align}
\alpha_{n}(\mathcal{V}) &= \left( \frac{2\epsilon^{2}}{1+\epsilon^{2}}\right)^{n}\int\mathrm{d}\nu\; \frac{(\nu-\mathcal{V})^{n}}{4(\phi+\phi^{-1})}
\notag\\
&\quad \times\left[\phi^{3}(\nu-\mathcal{V})\theta(\nu-\mathcal{V})e^{-{\epsilon^{2}\phi^{2}\nu^{2}}/{2}}\right.
\notag\\
&\qquad \left. +\phi^{-3}(\mathcal{V}-\nu)\theta(\mathcal{V}-\nu)e^{-{\epsilon^{2}\phi^{-2}\nu^{2}}/{2}}\right].
\end{align}
Performing a perturbation expansion in powers of $\epsilon$ for $\alpha_{n}(\mathcal{V})$ with
\begin{align}
\int_{0}^{\infty}\mathrm{d}\nu\; \nu^{n}e^{-{\epsilon^{2}\phi^{2}\nu^{2}}/{2}} &= \frac{\sqrt{2}^{n-1}\phi^{-n-1}}{\epsilon^{n+1}}\Gamma\left(\frac{n+1}{2}\right),
\\
\int_{-\infty}^{0}\mathrm{d}\nu\; \nu^{n}e^{-{\epsilon^{2}\phi^{-2}\nu^{2}}/{2}} &= \frac{(-1)^{n}\sqrt{2}^{n-1}\phi^{n+1}}{\epsilon^{n+1}}\Gamma\left(\frac{n+1}{2}\right),
\end{align}
where $\Gamma$ is the gamma function, we obtain
\begin{align}
\alpha_{1}(\mathcal{V}) &= -\mathcal{V}+\epsilon\eta\mathcal{V}^{2}+O(\epsilon^{2}),
\\
\alpha_{2}(\mathcal{V}) &= 2+O(\epsilon^{2}),
\\
\alpha_{3}(\mathcal{V}) &=-12\epsilon\eta+O(\epsilon^{2}),
\\
\alpha_{i}(\mathcal{V}) &= O(\epsilon^{2}), \;\; \text{for } i\geq 4,
\end{align}
where $\eta$ is a dimensionless parameter specifying the degree of nonequilibrium defined as
\begin{align}
\eta\equiv\sqrt{\frac{\pi}{8}}\left( \phi-\phi^{-1}\right) =\sqrt{\frac{\pi}{8}}\frac{\sqrt{T_{\mathrm{R}}}-\sqrt{T_{\mathrm{L}}}}{\sqrt{T}}.
\end{align}
Note that we do not assume $\eta\ll 1$.
We expand (\ref{eq:masterKMres}) in powers of $\epsilon$ as
\begin{align}
\frac{\partial \mathcal{P}(\mathcal{V},\tau)}{\partial \tau} &= \frac{\partial}{\partial\mathcal{V}}\left[\left( \mathcal{V}-\epsilon\eta\mathcal{V}^{2}\right) \mathcal{P}(\mathcal{V},\tau)\right]+\frac{\partial^{2}\mathcal{P}(\mathcal{V},\tau)}{\partial\mathcal{V}^{2}}
\notag\\
&\quad +2\epsilon\eta\frac{\partial^{3}\mathcal{P}(\mathcal{V},\tau)}{\partial\mathcal{V}^{3}}+O(\epsilon^{2}),
\label{eq:master_ep}
\end{align}
which contains only $\epsilon$ and $\eta$ as the dimensionless parameters at least to first order in $\epsilon$.
The first term on the right-hand side of this equation indicates that the relaxation time of $\mathcal{V}$ to lowest order in $\epsilon$ is set to unity by the rescaling (\ref{eq:res_vari}).
The escape rate for (\ref{eq:masterKMres}) is given by
\begin{align}
\alpha_{0}(\mathcal{V})&=\frac{1}{4\epsilon^{2}}-\frac{\eta\mathcal{V}}{2\epsilon}+O(\epsilon^{0}).
\end{align}
This implies that the average time between collisions is of order $\epsilon^{2}$ and much shorter than the relaxation time of $\mathcal{V}$.
Thus, for $\epsilon\ll 1$, we would naively expect that random collisions with the ideal gas particles are well approximated by a Gaussian white noise in a mesoscopic description due to the central limit theorem.
Hereafter, we ignore the $O(\epsilon^{2})$ terms in (\ref{eq:master_ep}).

\section{Preliminaries: the steady-state distribution and the moments}\label{sec:Pss}

Here, we review basic properties of the equation (\ref{eq:master_ep}).
Let $\mathcal{P}_{\mathrm{ss}}(\mathcal{V})$ denote the steady-state distribution of (\ref{eq:master_ep}).
From (\ref{eq:master_ep}), we have
\begin{align}
0=\left( \mathcal{V}-\epsilon\eta\mathcal{V}^{2}\right) \mathcal{P}_{\mathrm{ss}}(\mathcal{V})+\frac{\partial\mathcal{P}_{\mathrm{ss}}(\mathcal{V})}{\partial\mathcal{V}}+2\epsilon\eta\frac{\partial^{2}\mathcal{P}_{\mathrm{ss}}(\mathcal{V})}{\partial\mathcal{V}^{2}},
\label{eq:master_Pss}
\end{align}
where the left-hand side of this equation is zero because the probability current is zero at infinity.
Substituting
\begin{align}
\mathcal{P}_{\mathrm{ss}}(\mathcal{V}) = \frac{1}{\sqrt{2\pi}}e^{-\mathcal{V}^{2}/2}\Big[ 1+\epsilon f(\mathcal{V})\Big]
\end{align}
into (\ref{eq:master_Pss}) and ignoring $O(\epsilon^{2})$ terms, we get
\begin{equation}
\frac{\mathrm{d}f(\mathcal{V})}{\mathrm{d}\mathcal{V}}=2\eta-\eta\mathcal{V}^{2}.
\end{equation}
From the above equation with $\int\mathrm{d}\mathcal{V}\mathcal{P}_{\mathrm{ss}}(\mathcal{V})=1$, we obtain
\begin{align}
\mathcal{P}_{\mathrm{ss}}(\mathcal{V})=\frac{1}{\sqrt{2\pi}}e^{-\mathcal{V}^{2}/2}\left( 1+2\epsilon\eta\mathcal{V}-\frac{\epsilon\eta\mathcal{V}^{3}}{3} \right).
\label{eq:Pss}
\end{align}
Let $\langle\,\cdot\,\rangle_{\mathrm{ss}}$ denote the expectations with respect to $\mathcal{P}_{\mathrm{ss}}(\mathcal{V})$.
The moment- and cumulant-generating functions of $\mathcal{P}_{\mathrm{ss}}(\mathcal{V})$ are given by
\begin{align}
M(h)&\equiv\left\langle e^{h\mathcal{V}}\right\rangle_{\mathrm{ss}}=e^{h^{2}/2}\left( 1+\epsilon\eta h-\frac{\epsilon\eta h^{3}}{3}\right),
\label{eq:mgf_V}
\\
K(h)&\equiv \log\left\langle e^{h\mathcal{V}}\right\rangle_{\mathrm{ss}}=\epsilon\eta h+\frac{h^{2}}{2}-\frac{\epsilon\eta h^{3}}{3},
\label{eq:cgf_V}
\end{align}
to first order in $\epsilon$, respectively.
From (\ref{eq:mgf_V}), we get
\begin{align}
\langle\mathcal{V}\rangle_{\mathrm{ss}}=\epsilon\eta,\quad \langle\mathcal{V}^{2}\rangle_{\mathrm{ss}}=1,
\label{eq:aveV_varV}
\end{align}
which implies that $T$ is the kinetic temperature of the piston.
These results are consistent with the previous study~\cite{Gruber1999I}.
In (\ref{eq:Pss}), $\mathcal{P}_{\mathrm{ss}}(\mathcal{V})$ is negative for very large $\mathcal{V}$ when $\epsilon \ll 1$.
We, however, claim that such velocity will not be practically observed for $\epsilon\ll 1$ as the statistical expectations (\ref{eq:aveV_varV}) hold. 
To first order in $\epsilon (\ll 1)$, (\ref{eq:mgf_V}) and (\ref{eq:cgf_V}) are not affected because the negative probability of $\mathcal{P}_{\mathrm{ss}}(\mathcal{V})$ is less than $O(\exp [-(3/\epsilon\eta)^{2/3} / 2 ])$.
In Appendix \ref{sec:app_momeqs}, we provide another derivation of (\ref{eq:aveV_varV}) using relations between the moments of $\mathcal{P}_{\mathrm{ss}}(\mathcal{V})$.

\section{Large deviation function of time-averaged velocity}\label{sec:LD}

In this section, we analytically investigate statistical properties of the long-time averaged velocity
\begin{align}
\overline{\mathcal{V}}\equiv\frac{1}{\mathcal{T}}\int_{0}^{\mathcal{T}}\mathrm{d}\tau\; \mathcal{V}(\tau),
\label{eq:taveVori}
\end{align}
where $\mathcal{V}(\tau)$ is the velocity of the piston at time $\tau$.
Let $\mathcal{P}_{\mathcal{T}}(\overline{\mathcal{V}})$ denote the probability density of $\overline{\mathcal{V}}$.
The statistical properties of $\overline{\mathcal{V}}$ in the long-time limit are characterized by the large deviation function defined as
\begin{align}
\mathcal{I}(\overline{\mathcal{V}}) \equiv -\lim_{\mathcal{T}\to\infty}\frac{1}{\mathcal{T}}\log \mathcal{P}_{\mathcal{T}}(\overline{\mathcal{V}}).
\label{eq:I_def}
\end{align}
The statistical properties are also characterized by the scaled cumulant generating function defined as
\begin{align}
\mathcal{G}(h)\equiv\lim_{\mathcal{T}\to\infty}\frac{1}{\mathcal{T}}\log \int\mathrm{d}\overline{\mathcal{V}}\; e^{h\mathcal{T}\overline{\mathcal{V}}}\mathcal{P}_{\mathcal{T}}(\overline{\mathcal{V}}),
\label{eq:Gh_def}
\end{align}
which satisfies
\begin{align}
\mathcal{I}(\overline{\mathcal{V}})=\max_{h\in\mathbb{R}}\big[ h\overline{\mathcal{V}}-\mathcal{G}(h)\big].
\label{eq:I_G}
\end{align}
That is, $\mathcal{I}(\overline{\mathcal{V}})$ is the Legendre transform of $\mathcal{G}(h)$, which is known as the G\"{a}rtner--Ellis theorem in probability theory~\cite{Dembo1998,Touchette2009}.
Because $\mathcal{G}(h)$ equals the dominant eigenvalue of some linear operator~\cite{Garrahan2009,Nemoto2011II,Touchette2018}, $\mathcal{G}(h)$ is easier to calculate than $\mathcal{I}(\overline{\mathcal{V}})$.

We next identify the linear operator whose dominant eigenvalue is equal to $\mathcal{G}(h)$.
Let $\mathcal{X}(\tau)$ and $\mathcal{P}(\mathcal{X},\mathcal{V},\tau)$ denote the displacement of the piston over a time interval $[0,\tau]$ and the joint probability density of $\mathcal{X}({\tau})$ and $\mathcal{V}(\tau)$.
Note that $\mathcal{X}(\mathcal{T})=\mathcal{T}\overline{\mathcal{V}}$ and
\begin{align}
\int\mathrm{d}\overline{\mathcal{V}}\; e^{h\mathcal{T}\overline{\mathcal{V}}}\mathcal{P}_{\mathcal{T}}(\overline{\mathcal{V}}) = \int\mathrm{d}\mathcal{X}\int\mathrm{d}\mathcal{V}\; e^{h\mathcal{X}}\mathcal{P}(\mathcal{X},\mathcal{V},\mathcal{T}).
\label{eq:PXVPT}
\end{align}
Using $\partial_{\tau}\mathcal{X}({\tau})=\mathcal{V}(\tau)$ and (\ref{eq:master_ep}), we have
\begin{align}
\frac{\partial \mathcal{P}(\mathcal{X},\mathcal{V},\tau)}{\partial \tau} &= -\frac{\partial}{\partial\mathcal{X}}\Big[ \mathcal{V}\mathcal{P}(\mathcal{X},\mathcal{V},\tau)\Big]
\notag\\[3pt]
&\quad +\frac{\partial}{\partial\mathcal{V}}\left[\left( \mathcal{V}-\epsilon\eta\mathcal{V}^{2}\right) \mathcal{P}(\mathcal{X},\mathcal{V},\tau)\right]
\notag\\[3pt]
&\quad +\frac{\partial^{2}\mathcal{P}(\mathcal{X},\mathcal{V},\tau)}{\partial\mathcal{V}^{2}}+2\epsilon\eta\frac{\partial^{3}\mathcal{P}(\mathcal{X},\mathcal{V},\tau)}{\partial\mathcal{V}^{3}}.
\label{eq:master_XV}
\end{align}
Introducing a quantity
\begin{align}
\mathcal{Q}_{h}(\mathcal{V},\tau)\equiv\int\mathrm{d}\mathcal{X}\; e^{h\mathcal{X}}\mathcal{P}(\mathcal{X},\mathcal{V},\tau),
\label{eq:def_Q}
\end{align}
and using (\ref{eq:master_XV}), we get
\begin{align}
\frac{\partial \mathcal{Q}_{h}(\mathcal{V},\tau)}{\partial \tau}&=\mathcal{L}_{h}^{\dagger}\mathcal{Q}_{h}(\mathcal{V},\tau)
\label{eq:evo_Q}
\end{align}
with the operator
\begin{align}
\mathcal{L}_{h}^{\dagger}&=h\mathcal{V}+\frac{\partial}{\partial\mathcal{V}}\left( \mathcal{V}-\epsilon\eta\mathcal{V}^{2}\right)+\frac{\partial^{2}}{\partial\mathcal{V}^{2}}+2\epsilon\eta\frac{\partial^{3}}{\partial\mathcal{V}^{3}}.
\end{align}
Let $\mu_{h}$ and $\Phi_{h}(\mathcal{V})$ denote the dominant eigenvalue and corresponding right eigenfunction of $\mathcal{L}_{h}^{\dagger}$. 
Because $\mathcal{Q}_{h}(\mathcal{V},\tau) > 0$ for any $\tau$ from (\ref{eq:def_Q}), we expect that $\mu_{h}$ is real and $\Phi_{h}(\mathcal{V})>0$.
We perturbatively check these conditions later.
Then, $\mathcal{Q}_{h}(\mathcal{V},\mathcal{T})$ is dominated by the term associated with the largest eigenvalue for sufficiently large $\mathcal{T}$.
Thus, from (\ref{eq:evo_Q}), $\mathcal{Q}_{h}(\mathcal{V},\mathcal{T})$ is asymptotically given by
\begin{align}
\mathcal{Q}_{h}(\mathcal{V},\mathcal{T}) \simeq  c_0 \Phi_{h}(\mathcal{V}) e^{\mu_{h}\mathcal{T}},
\end{align}
where the constant $c_{0}$ is determined by the initial condition $\mathcal{Q}_{h}(\mathcal{V},0)$.
Using the above equation, we get
\begin{align}
\lim_{\mathcal{T}\to\infty}\frac{1}{\mathcal{T}}\log\int\mathrm{d}\mathcal{V}\; \mathcal{Q}_{h}(\mathcal{V},\mathcal{T}) = \mu_{h}.
\end{align}
From this equation, (\ref{eq:Gh_def}), (\ref{eq:PXVPT}), and (\ref{eq:def_Q}), we obtain
\begin{align}
\mathcal{G}(h)=\mu_{h},
\label{eq:mu_G}
\end{align}
which shows that the scaled cumulant generating function, $\mathcal{G}(h)$, is given by the largest eigenvalue of $\mathcal{L}_{h}^{\dagger}$.

For later convenience, for any smooth functions $\mathcal{Y}$ and $\mathcal{Z}$, we define an inner product by
\begin{align}
\left\langle \mathcal{Y},\mathcal{Z}\right\rangle\equiv\int\mathrm{d}\mathcal{V}\; \mathcal{Y}(\mathcal{V})\mathcal{Z}(\mathcal{V}).
\end{align}
Then, the adjoint operator of $\mathcal{L}_{h}^{\dagger}$ denoted by $\mathcal{L}_{h}$ is given by
\begin{align}
\langle \mathcal{Y},\mathcal{L}_{h}^{\dagger}\mathcal{Z}\rangle = \langle \mathcal{L}_{h}\mathcal{Y},\mathcal{Z}\rangle.
\end{align}
The adjoint operator $\mathcal{L}_{h}$ is explicitly written as
\begin{align}
\mathcal{L}_{h}&=h\mathcal{V}-\left( \mathcal{V}-\epsilon\eta\mathcal{V}^{2}\right) \frac{\partial}{\partial\mathcal{V}}+\frac{\partial^{2}}{\partial\mathcal{V}^{2}}-2\epsilon\eta\frac{\partial^{3}}{\partial\mathcal{V}^{3}}.
\end{align}
Because $\mu_{h}$ is real, the largest eigenvalue of $\mathcal{L}_{h}$ equals $\mu_{h}$.
Let $\Psi_{h}(\mathcal{V})$ denote the largest right eigenfunction of $\mathcal{L}_{h}$.

We now expand $\mathcal{L}_{h}$, $\mu_{h}$, $\Psi_{h}$, and $\Phi_{h}$ in powers of $\epsilon$:
\begin{align}
\mathcal{L}_{h}&=\mathcal{L}_{h}^{(0)}+\epsilon\mathcal{L}_{h}^{(1)},
\\
\mu_{h}&=\mu_{h}^{(0)}+\epsilon\mu_{h}^{(1)},
\\
\Psi_{h}&=\Psi_{h}^{(0)}+\epsilon\Psi_{h}^{(1)},
\\
\Phi_{h}&=\Phi_{h}^{(0)}+\epsilon\Phi_{h}^{(1)},
\end{align}
and perturbatively calculate $\mu_{h}$.
Note that $\mathcal{L}_{h}^{(0)}$ and $\mathcal{L}_{h}^{(1)}$ are explicitly given by
\begin{align}
\mathcal{L}_{h}^{(0)}&=h\mathcal{V}-\mathcal{V}\frac{\partial}{\partial\mathcal{V}}+\frac{\partial^{2}}{\partial\mathcal{V}^{2}},
\\
\mathcal{L}_{h}^{(1)}&=\eta\mathcal{V}^{2}\frac{\partial}{\partial\mathcal{V}}-2\eta\frac{\partial^{3}}{\partial\mathcal{V}^{3}}.
\end{align}

The eigenvalue equation to zeroth order in $\epsilon$ is given by
\begin{align}
\mathcal{L}_{h}^{(0)}\Psi_{h}^{(0)}=\mu_{h}^{(0)}\Psi_{h}^{(0)}.
\label{eq:ee_0}
\end{align}
Using symmetrization, the largest eigenvalue of $\mathcal{L}_{h}^{(0)}$ is obtained by calculating ground state energy of some quantum system~\cite{Majumdar2002,Touchette2018}.
Substituting
\begin{align}
\Psi_{h}^{(0)}=e^{\frac{\mathcal{V}^{2}}{4}}\psi_{h}^{(0)}
\end{align}
into (\ref{eq:ee_0}), we get
\begin{align}
-\frac{\partial^{2}\psi_{h}^{(0)}}{\partial\mathcal{V}^{2}} +\left(\frac{\mathcal{V}}{2}-h\right)^{2}\psi_{h}^{(0)}=\left( h^{2}-\mu_{h}^{(0)}+\frac{1}{2}\right) \psi_{h}^{(0)},
\end{align}
which is equivalent to the one-dimensional Schr\"odinger equation for the harmonic oscillator.
Thus, we obtain
\begin{align}
\mu_{h}^{(0)}=h^{2},\quad \psi_{h}^{(0)}=e^{-\frac{\mathcal{V}^{2}}{4}+h\mathcal{V}}, \quad \Psi_{h}^{(0)}=e^{h\mathcal{V}}.
\label{eq:mu_0}
\end{align}
Similarly, we can also calculate $\Phi_{h}^{(0)}$ as
\begin{align}
\Phi_{h}^{(0)}&=e^{-\frac{\mathcal{V}^{2}}{2}+h\mathcal{V}},
\end{align}
and confirm that $\mu_{h}^{(0)}$ is real and $\Phi_{h}^{(0)}>0$.

The eigenvalue equation to first order in $\epsilon$ is given by
\begin{align}
\mathcal{L}_{h}^{(0)}\Psi_{h}^{(1)}+\mathcal{L}_{h}^{(1)}\Psi_{h}^{(0)}=\mu_{h}^{(1)}\Psi_{h}^{(0)}+\mu_{h}^{(0)}\Psi_{h}^{(1)}.
\label{eq:ee_1}
\end{align}
The solvability condition for (\ref{eq:ee_1}) leads to
\begin{align}
& \left\langle \Phi_{h}^{(0)},\mathcal{L}_{h}^{(0)}\Psi_{h}^{(1)}\right\rangle + \left\langle \Phi_{h}^{(0)},\mathcal{L}_{h}^{(1)}\Psi_{h}^{(0)}\right\rangle
\notag\\
& \quad = \mu_{h}^{(1)}\left\langle \Phi_{h}^{(0)},\Psi_{h}^{(0)}\right\rangle + \mu_{h}^{(0)}\left\langle \Phi_{h}^{(0)},\Psi_{h}^{(1)}\right\rangle.
\end{align}
Using
\begin{align}
\left\langle \Phi_{h}^{(0)},\mathcal{L}_{h}^{(0)}\Psi_{h}^{(1)}\right\rangle=\mu_{h}^{(0)}\left\langle \Phi_{h}^{(0)},\Psi_{h}^{(1)}\right\rangle,
\end{align}
we have
\begin{align}
\mu_{h}^{(1)}&=\frac{\left\langle \Phi_{h}^{(0)},\mathcal{L}_{h}^{(1)}\Psi_{h}^{(0)}\right\rangle}{\left\langle \Phi_{h}^{(0)},\Psi_{h}^{(0)}\right\rangle}=\eta h+2\eta h^{3},
\label{eq:mu_1}
\end{align}
and confirm that $\mu_{h}^{(1)}$ is real.
Substituting (\ref{eq:mu_0}) and (\ref{eq:mu_1}) into (\ref{eq:ee_1}) and considering the condition $\langle \Phi_{h}^{(0)},\Psi_{h}^{(1)}\rangle < \infty$, we have
\begin{align}
\Psi_{h}^{(1)}&=\eta\left( 2h^{2}\mathcal{V}+\frac{h\mathcal{V}^{2}}{2}+c\right) e^{h\mathcal{V}},
\end{align}
where $c$ is an arbitrary constant.
Repeating this procedure, we may obtain higher order terms in $\epsilon$.
See Appendix~\ref{sec:app_second} for second-order calculations.

From (\ref{eq:I_G}) and (\ref{eq:mu_G}), we finally obtain to first order in $\epsilon$
\begin{align}
\mathcal{G}(h) &= \epsilon\eta h+h^{2}+2\epsilon\eta h^{3},
\label{eq:Gh}
\\[3pt]
\mathcal{I}(\overline{\mathcal{V}}) &=\frac{\overline{\mathcal{V}}^{2}}{4}-\epsilon\eta\left( \frac{\overline{\mathcal{V}}}{2}+\frac{\overline{\mathcal{V}}^{3}}{4}\right).
\label{eq:I}
\end{align}
The result (\ref{eq:Gh}) is consistent with that obtained numerically in~\cite{Seya2020}.
$\mathcal{G}(h)$ in (\ref{eq:Gh}) is not convex on the interval $(-\infty,-1/(6\epsilon\eta))$ because we have ignored the contribution of $O(\epsilon^2)$ to derive (\ref{eq:Gh}).

\section{Effective Langevin equations}\label{sec:ELE}

In this section, we investigate effective models for the nonequilibrium Rayleigh piston.
Putting aside the derivation of the effective models from the microscopic description, we first seek the models that reproduce the large deviation function (\ref{eq:I}) as a necessary condition.
We show that there exists an infinite number of Langevin equations with the same large deviation (\ref{eq:I}).
Let us introduce the following Langevin equation for the non-dimensionalized velocity $\hat{\mathcal{V}}(\tau)$:
\begin{align}
\frac{\mathrm{d}\hat{\mathcal{V}}(\tau)}{\mathrm{d}\tau}&=-\hat{\mathcal{V}}(\tau)+\epsilon \left( \mathcal{A}_{0}+\mathcal{A}_{1}\hat{\mathcal{V}}(\tau)+\mathcal{A}_{2}\hat{\mathcal{V}}^{2}(\tau)\right) 
\notag\\
&\quad +\sqrt{2+2\epsilon\left( \mathcal{B}_{0}+\mathcal{B}_{1}\hat{\mathcal{V}}(\tau)\right)}\cdot\xi(\tau),
\label{eq:LE}
\end{align}
where $\mathcal{A}_{0}$, $\mathcal{A}_{1}$, $\mathcal{A}_{2}$, $\mathcal{B}_{0}$, and $\mathcal{B}_{1}$ are $\epsilon$-independent constants, $\xi$ is a zero-mean Gaussian white noise with unit variance $\langle\xi(\tau)\xi(\tau')\rangle=\delta(\tau-\tau')$, and the symbol $\cdot$ denotes the It\^o-product.
The hat symbol indicates quantities associated with the Langevin equation (\ref{eq:LE}).
The noise intensity of (\ref{eq:LE}) is expected to be positive for $\epsilon\ll 1$.
Note that the relation between the velocity of the piston at time $\tau$ in the original model, $\mathcal{V}(\tau)$, and $\hat{\mathcal{V}}(\tau)$ is unclear although $\hat{\mathcal{V}}(\tau)$ may be related to an effective velocity obtained by coarse-graining $\mathcal{V}(\tau)$ due to the central limit theorem.
Let $\hat{\mathcal{P}}(\hat{\mathcal{V}},\tau)$ denote the probability density of $\hat{\mathcal{V}}(\tau)$.
The corresponding Fokker--Planck equation is given by
\begin{align}
\frac{\partial \hat{\mathcal{P}}(\hat{\mathcal{V}},\tau)}{\partial \tau} &= \frac{\partial}{\partial\hat{\mathcal{V}}}\left[ \hat{\mathcal{V}}-\epsilon\left( \mathcal{A}_{0}+\mathcal{A}_{1}\hat{\mathcal{V}}+\mathcal{A}_{2}\hat{\mathcal{V}}^{2}\right) \right] \hat{\mathcal{P}}(\hat{\mathcal{V}},\tau)
\notag\\
&\quad +\frac{\partial^{2}}{\partial\hat{\mathcal{V}}^{2}}\left[1+\epsilon\left( \mathcal{B}_{0}+\mathcal{B}_{1}\hat{\mathcal{V}}\right)\right] \hat{\mathcal{P}}(\hat{\mathcal{V}},\tau).
\label{eq:FP}
\end{align}

Following the same procedure as in Sec.~\ref{sec:LD}, we calculate the scaled cumulant generating function, $\hat{\mathcal{G}}(h)$, of the time-averaged velocity for (\ref{eq:LE}).
Using
\begin{align}
\hat{\mathcal{L}}_{h}^{(1)}=\left( \mathcal{A}_{0}+\mathcal{A}_{1}\hat{\mathcal{V}}+\mathcal{A}_{2}\hat{\mathcal{V}}^{2}\right) \frac{\partial}{\partial\hat{\mathcal{V}}}+\left( \mathcal{B}_{0}+\mathcal{B}_{1}\hat{\mathcal{V}}\right)\frac{\partial^{2}}{\partial\hat{\mathcal{V}}^{2}},
\end{align}
instead of $\mathcal{L}_{h}^{(1)}$, we get
\begin{align}
\frac{\left\langle \Phi_{h}^{(0)},\hat{\mathcal{L}}_{h}^{(1)}\Psi_{h}^{(0)}\right\rangle}{\left\langle \Phi_{h}^{(0)},\Psi_{h}^{(0)}\right\rangle}&=\left( \mathcal{A}_{0}+\mathcal{A}_{2}\right) h +\left( 2\mathcal{A}_{1}+\mathcal{B}_{0}\right) h^{2}
\notag\\
&\quad +\left( 4\mathcal{A}_{2}+2\mathcal{B}_{1}\right) h^{3},
\label{eq:hatmu_1}
\end{align}
which leads to
\begin{align}
\hat{\mathcal{G}}(h)&=\epsilon \left( \mathcal{A}_{0}+\mathcal{A}_{2}\right) h + \big[ 1+\epsilon\left( 2\mathcal{A}_{1}+\mathcal{B}_{0}\right)\big] h^{2}
\notag\\
&\quad +\epsilon\left( 4\mathcal{A}_{2}+2\mathcal{B}_{1}\right) h^{3}+O(\epsilon^{2}).
\label{eq:hatGh}
\end{align}
From (\ref{eq:Gh}) and (\ref{eq:hatGh}), we have $\hat{\mathcal{G}}(h)=\mathcal{G}(h)$ by setting 
\begin{align}
\begin{cases}
\mathcal{A}_{0}=\eta-\mathcal{A}_{2},
\\
\mathcal{B}_{0}=-2\mathcal{A}_{1},
\\
\mathcal{B}_{1}=\eta-2\mathcal{A}_{2}.
\end{cases}
\label{eq:Gh_cond}
\end{align}
Thus, the Langevin equation
\begin{align}
\frac{\mathrm{d}\hat{\mathcal{V}}(\tau)}{\mathrm{d}\tau}&=\epsilon\left(\eta-\mathcal{A}_{2}\right)-\left(1-\epsilon\mathcal{A}_{1}\right)\hat{\mathcal{V}}(\tau)+\epsilon\mathcal{A}_{2}\hat{\mathcal{V}}^{2}(\tau)
\notag\\
&\quad +\sqrt{2-4\epsilon\mathcal{A}_{1}+2\epsilon\left(\eta-2\mathcal{A}_{2}\right)\hat{\mathcal{V}}(\tau)}\cdot\xi(\tau),
\label{eq:ELE}
\end{align}
gives the same large deviation function as in (\ref{eq:I}) for any $\mathcal{A}_{1}$ and $\mathcal{A}_{2}$ to first order in $\epsilon$.
The result (\ref{eq:ELE}) implies that the effective Langevin equation for (\ref{eq:master_ep}) is not uniquely determined by only the large deviation function of the time-averaged velocity.

Next, we investigate what conditions can be used in addition to $\hat{\mathcal{G}}(h)=\mathcal{G}(h)$ to determine the form of the Langevin equation (\ref{eq:ELE}).
As one attempt to determine $\mathcal{A}_{1}$ and $\mathcal{A}_{2}$ in (\ref{eq:ELE}), we derive the Langevin equation with the same cumulant-generating function as in (\ref{eq:cgf_V}).
Using the same method as in Sec.~\ref{sec:Pss}, the steady-state distribution of (\ref{eq:FP}) with (\ref{eq:Gh_cond}) is given by
\begin{align}
\hat{\mathcal{P}}_{\mathrm{ss}}(\hat{\mathcal{V}})&=\frac{1}{\sqrt{2\pi}}e^{-\hat{\mathcal{V}}^{2}/2}\left[ 1+\frac{\epsilon\mathcal{A}_{1}}{2}+\epsilon\mathcal{A}_{2}\hat{\mathcal{V}}\right.
\notag\\
&\quad \left. -\frac{\epsilon\mathcal{A}_{1}\hat{\mathcal{V}}^{2}}{2}+\frac{\epsilon\left(\eta-\mathcal{A}_{2}\right)\hat{\mathcal{V}}^{3}}{3}\right] +O(\epsilon^{2}).
\label{eq:FP_Pss}
\end{align}
Note that there may be no steady-state distribution for (\ref{eq:LE}) with some specific parameters. We later check the existence of the steady state distribution.
The moment- and cumulant-generating functions of the distribution (\ref{eq:FP_Pss}) are given by
\begin{align}
\hat{M}(h)&=e^{h^{2}/2}\bigg[ 1+\epsilon\eta h-\frac{\epsilon\mathcal{A}_{1}h^{2}}{2}+\frac{\epsilon\left(\eta-\mathcal{A}_{2}\right)h^{3}}{3}\bigg],
\\[3pt]
\hat{K}(h)&=\epsilon\eta h+\frac{\left( 1-\epsilon\mathcal{A}_{1}\right) h^{2}}{2} +\frac{\epsilon\left(\eta-\mathcal{A}_{2}\right)h^{3}}{3},
\label{eq:hatcgf_V}
\end{align}
where we have ignored $O(\epsilon^{2})$ terms.
Comparing (\ref{eq:cgf_V}) and (\ref{eq:hatcgf_V}), we obtain $\hat{K}(h)=K(h)$ by setting 
\begin{align}
\begin{cases}
\mathcal{A}_{1}=0,
\\
\mathcal{A}_{2}=2\eta.
\end{cases}
\label{eq:cgf_cond}
\end{align}
From (\ref{eq:ELE}) with (\ref{eq:cgf_cond}), we obtain the Langevin equation
\begin{align}
\frac{\mathrm{d}\hat{\mathcal{V}}(\tau)}{\mathrm{d}\tau}&=-\epsilon\eta-\hat{\mathcal{V}}(\tau)+2\epsilon\eta\hat{\mathcal{V}}^{2}(\tau)
\notag\\
&\quad +\sqrt{2\left( 1-3\epsilon\eta\hat{\mathcal{V}}(\tau)\right)}\cdot\xi(\tau).
\label{eq:ELE_CGF}
\end{align}
We can derive the exact steady-state distribution for (\ref{eq:ELE_CGF}), and the noise intensity of (\ref{eq:ELE_CGF}) is always positive even for $\epsilon \nll 1$.
See Appendix \ref{sec:app_NI} for the details.
Note that the existence of the upper bound on $\hat{\mathcal{V}}(\tau)$ is our computational artifact because the original model has no bound for the velocity $\mathcal{V}(\tau)$.
Because the upper bound on $\hat{\mathcal{V}}(\tau)$ is given by $1/(3\epsilon\eta)$, we may ignore the effects of the bound when $\epsilon\ll 1$.
Here, we define the time averaged velocity for (\ref{eq:master_ep}) and (\ref{eq:ELE_CGF}) in the steady state by
\begin{align}
\overline{\mathcal{V}}(\tau) &\equiv\frac{1}{\mathcal{T}}\int_{\tau}^{\tau+\mathcal{T}}\mathrm{d}\tau'\; \mathcal{V}(\tau'),
\label{eq:t_ave_V}
\\
\overline{\hat{\mathcal{V}}}(\tau) &\equiv \frac{1}{\mathcal{T}}\int_{\tau}^{\tau+\mathcal{T}}\mathrm{d}\tau'\; \hat{\mathcal{V}}(\tau'),
\end{align}
respectively, for $\tau\gg 1$.
Note that $\overline{\mathcal{V}}(0)=\overline{\mathcal{V}}$, where $\overline{\mathcal{V}}$ is defined in (\ref{eq:taveVori}).
The statistical property of $\overline{\hat{\mathcal{V}}}(\tau)$ is equal to that of $\overline{\mathcal{V}}(\tau)$ when $\mathcal{T}\ll 1$ and $\mathcal{T}\gg 1$.

As another attempt to determine $\mathcal{A}_{1}$ and $\mathcal{A}_{2}$ in (\ref{eq:ELE}), we assume that the drift coefficient of (\ref{eq:master_ep}) equals that of (\ref{eq:FP}) with (\ref{eq:Gh_cond}) instead of $\hat{K}(h)=K(h)$.
This assumption leads to
\begin{align}
\begin{cases}
\mathcal{A}_{1}=0,
\\
\mathcal{A}_{2}=\eta.
\end{cases}
\label{eq:dc_cond}
\end{align}
From (\ref{eq:ELE}) with (\ref{eq:dc_cond}), we obtain the Langevin equation
\begin{align}
\frac{\mathrm{d}\hat{\mathcal{V}}(\tau)}{\mathrm{d}\tau}&=-\left( 1-\epsilon\eta\hat{\mathcal{V}}(\tau)\right) \hat{\mathcal{V}}(\tau) +\sqrt{2\left( 1-\epsilon\eta\hat{\mathcal{V}}(\tau)\right)}\cdot\xi(\tau),
\notag\\
&= \epsilon\eta -\left( 1-\epsilon\eta\hat{\mathcal{V}}(\tau)\right) \hat{\mathcal{V}}(\tau) 
\notag\\
& \qquad +\sqrt{2\left( 1-\epsilon\eta\hat{\mathcal{V}}(\tau)\right)}\odot\xi(\tau),
\label{eq:ELE_DC}
\end{align}
where the symbol $\odot$ denotes the anti-It\^o product.
This equation was numerically found in~\cite{Seya2020}.
In this form, a fluctuation-dissipation-like relation holds even out of equilibrium.

Note that when we use instead of (\ref{eq:LE})
\begin{align}
&\frac{\mathrm{d}\hat{\mathcal{V}}(\tau)}{\mathrm{d}\tau}=-\hat{\mathcal{V}}(\tau)
\notag\\
& \qquad +\epsilon \left( \mathcal{A}_{0}+\mathcal{A}_{1}\hat{\mathcal{V}}(\tau)+\mathcal{A}_{2}\hat{\mathcal{V}}^{2}(\tau)+\mathcal{A}_{3}\hat{\mathcal{V}}^{3}(\tau)+\cdots\right) 
\notag\\
& \qquad +\sqrt{2+2\epsilon\left( \mathcal{B}_{0}+\mathcal{B}_{1}\hat{\mathcal{V}}(\tau)+\mathcal{B}_{2}\hat{\mathcal{V}}^{2}(\tau)+\cdots\right)}\cdot\xi(\tau),
\label{eq:LEmore}
\end{align}
we cannot uniquely determine all coefficients by using the conditions discussed in this section.
We do not yet know how to determine all the coefficients.

\section{Concluding remarks}\label{sec:remarks}

We calculated the scaled cumulant generating function (\ref{eq:Gh}) and the large deviation function (\ref{eq:I}) of the time-averaged velocity for the nonequilibrium Rayleigh piston beyond the linear response regime.
The key point of the calculation is that the scaled cumulant generating function equals the largest eigenvalue of $\mathcal{L}_{h}$ and that the eigenvalue equation for $\mathcal{L}_{h}$ to zeroth order in $\epsilon$ is converted to the one-dimensional Schr\"odinger equation for the harmonic oscillator.
We identify effective Langevin equations (\ref{eq:ELE}) for the piston that reproduce the same large deviation function (\ref{eq:I}).
We provide two possibilities for uniquely determining the form of the effective model (\ref{eq:ELE}).
Before ending this paper, we will make some remarks.

We can impose further constraints on the Langevin equation (\ref{eq:ELE}) by assuming that (\ref{eq:ELE}) satisfies the same symmetry properties as (\ref{eq:master_ep}) to first order in $\epsilon$.
For example, the operator appeared in (\ref{eq:master_ep})
\begin{align}
\mathcal{L}^{\dagger}_{0}\equiv \frac{\partial}{\partial\mathcal{V}}\left( \mathcal{V}-\epsilon\eta\mathcal{V}^{2}\right)+\frac{\partial^{2}}{\partial\mathcal{V}^{2}}+2\epsilon\eta\frac{\partial^{3}}{\partial\mathcal{V}^{3}},
\end{align}
is invariant under the transformations $(\mathcal{V},\eta) \to (-\mathcal{V},-\eta)$ and $(\epsilon,\eta) \to (q\epsilon,\eta/q)$, where $q$ is a non-zero constant.
Thus, by imposing invariance on (\ref{eq:LE}) under these two transformations, we expect that $\mathcal{A}_{0}$, $\mathcal{A}_{2}$, and $\mathcal{B}_{1}$ is linear in $\eta$ and that $\mathcal{A}_{1}=\mathcal{B}_{0}=0$.
Using these conditions with (\ref{eq:ELE}), we get
\begin{align}
\frac{\mathrm{d}\hat{\mathcal{V}}(\tau)}{\mathrm{d}\tau}&=\epsilon\eta\left(1-\tilde{\mathcal{A}_{2}}\right)-\hat{\mathcal{V}}(\tau)+\epsilon\eta\tilde{\mathcal{A}}_{2}\hat{\mathcal{V}}^{2}(\tau)
\notag\\
&\quad +\sqrt{2+2\epsilon\eta\left(1-2\tilde{\mathcal{A}}_{2}\right)\hat{\mathcal{V}}(\tau)}\cdot\xi(\tau),
\label{eq:ELE_sym}
\end{align}
where $\tilde{\mathcal{A}}_{2}$ is an $\eta$-independent constant.
Considering the restriction in terms of symmetry properties is useful for determing the form of the possible Langevin equations.
We note that the Langevin equations (\ref{eq:ELE_CGF}) and (\ref{eq:ELE_DC}) satisfy the symmetry properties because the conditions used in addition to $\hat{\mathcal{G}}(h)=\mathcal{G}(h)$ to determine the form of the Langevin equation (\ref{eq:ELE}) do not break the symmetry properties.

By recalling that the average time between collisions is given by $4\epsilon^{2}$ to the lowest order in $\epsilon$, the stochastic evolution of the long-time-averaged velocity $\overline{\mathcal{V}}(\tau)$ with $4\epsilon^{2}\ll\mathcal{T}\ll 1$ should be effectively described by some Langevin equation due to the central limit theorem.
However, it is not yet known how to derive the Langevin equation for $\overline{\mathcal{V}}(\tau)$.
There is a possibility that the Langevin equation for $\overline{\mathcal{V}}(\tau)$ is not given by (\ref{eq:ELE}) with specific $\mathcal{A}_{1}$ and $\mathcal{A}_{2}$.
The substantive future study will be to derive numerically or analytically the Langevin equation from the microscopic description.
We believe that the analysis developed in this paper may be a first step towards an understanding of the universal description of the stochastic evolution of slow variables out of equilibrium.

\begin{acknowledgments}
The authors thank Takahiro Nemoto and Atsumasa Seya for useful comments.
This work was supported by JSPS KAKENHI Grant Numbers JP17H01148, JP17K14355, JP19H01864, JP19H05795, JP19K03647, JP20J00003, JP20K20425.
\end{acknowledgments}

\appendix
\onecolumngrid
\section{Relations between the moments}\label{sec:app_momeqs}

Multiplying (\ref{eq:master_Pss}) by $\mathcal{V}^{n-1}$ and integrating with $\mathcal{V}$, we have for $n\in\mathbb{Z}^{+}$
\begin{align}
\langle\mathcal{V}^{n}\rangle_{\mathrm{ss}}&=(n-1)\langle\mathcal{V}^{n-2}\rangle_{\mathrm{ss}}+\epsilon\eta\langle\mathcal{V}^{n+1}\rangle_{\mathrm{ss}} -2(n-1)(n-2)\epsilon\eta\langle\mathcal{V}^{n-3}\rangle_{\mathrm{ss}}.
\label{eq:moment_ss}
\end{align}
Using (\ref{eq:moment_ss}) with $\langle\mathcal{V}\rangle_{\mathrm{ss}}=\epsilon\eta\langle\mathcal{V}^{2}\rangle_{\mathrm{ss}}$, we get $\langle\mathcal{V}^{2n-1}\rangle_{\mathrm{ss}}=O(\epsilon)$.
Thus, we obtain
\begin{align}
\begin{cases}
\langle\mathcal{V}^{2n}\rangle_{\mathrm{ss}}=(2n-1)\langle\mathcal{V}^{2n-2}\rangle_{\mathrm{ss}},
\\[3pt]
\langle\mathcal{V}^{2n+1}\rangle_{\mathrm{ss}}=2n\langle\mathcal{V}^{2n-1}\rangle_{\mathrm{ss}}-(2n-1)\epsilon\eta\langle\mathcal{V}^{2n}\rangle_{\mathrm{ss}},
\end{cases}
\label{eq:moment_rel}
\end{align}
where we have ignored $O(\epsilon^{2})$ terms.
From (\ref{eq:moment_rel}) with $n=1$, we have $\langle\mathcal{V}^{2}\rangle_{\mathrm{ss}}=1$, which leads to $\langle\mathcal{V}\rangle_{\mathrm{ss}}=\epsilon\eta$.
Repeatedly using (\ref{eq:moment_rel}), we can calculate the $n$-th moment of $\mathcal{P}_{\mathrm{ss}}(\mathcal{V})$.

\section{Second-order calculations}\label{sec:app_second}

Using the same method as in Sec.~\ref{sec:model}, we further expand (\ref{eq:masterKMres}) in powers of $\epsilon$ as
\begin{align}
\frac{\partial \mathcal{P}(\mathcal{V},\tau)}{\partial \tau} &= \frac{\partial}{\partial\mathcal{V}}\left( \mathcal{V}-\epsilon\eta\mathcal{V}^{2}-\epsilon^{2}\mathcal{V}+\frac{\epsilon^{2}\left( 8\eta^{2}+\pi\right)\mathcal{V}^{3}}{6\pi}\right) \mathcal{P}(\mathcal{V},\tau)+\frac{\partial^{2}}{\partial\mathcal{V}^{2}}\left(1-2\epsilon^{2}+\frac{3\epsilon^{2}\mathcal{V}^{2}}{2}\right)\mathcal{P}(\mathcal{V},\tau)
\notag\\[3pt]
&\quad +\frac{\partial^{3}}{\partial\mathcal{V}^{3}}\left( 2\epsilon\eta+\frac{8\epsilon^{2}\mathcal{V}}{3}\right) \mathcal{P}(\mathcal{V},\tau) +\frac{4\epsilon^{2}\left( 8\eta^{2}+\pi\right)}{3\pi}\frac{\partial^{4}\mathcal{P}(\mathcal{V},\tau)}{\partial\mathcal{V}^{4}}+O(\epsilon^{3}).
\label{eq:master_ep2}
\end{align}
The steady-state distribution of this equation is given by
\begin{align}
\mathcal{P}_{\mathrm{ss}}(\mathcal{V})=\frac{1}{\sqrt{2\pi}}e^{-\mathcal{V}^{2}/2}\left[ 1+\epsilon\eta\left( 2\mathcal{V}-\frac{\mathcal{V}^{3}}{3}\right)+\epsilon^{2}\eta^{2}\left(\frac{54-23\pi}{6\pi}-\frac{16-8\pi}{\pi}\mathcal{V}^{2}+\frac{7-5\pi}{3\pi}\mathcal{V}^{4}+\frac{\mathcal{V}^{6}}{18}\right) +O(\epsilon^{3})\right],
\label{eq:Pss_ep2}
\end{align}
which is consistent with the result in~\cite{Gruber1999I}.
The moment- and cumulant-generating functions of $\mathcal{P}_{\mathrm{ss}}(\mathcal{V})$ are calculated as
\begin{align}
M(h)&=e^{h^{2}/2}\left[ 1+\epsilon\eta\left( h-\frac{h^{3}}{3}\right)+\epsilon^{2}\eta^{2}\left( \frac{\pi-4}{2\pi}h^{2}+\frac{14-5\pi}{6\pi}h^{4}+\frac{h^{6}}{18}\right)+O(\epsilon^{3})\right],
\label{eq:mgf_V_ep2}
\\[3pt]
K(h)&=\epsilon\eta h+\left(\frac{1}{2}-\frac{2\epsilon^{2}\eta^{2}}{\pi}\right)h^{2}-\frac{\epsilon\eta h^{3}}{3}+\frac{\left( 14-3\pi\right)\epsilon^{2}\eta^{2}h^{4}}{6\pi}+O(\epsilon^{3}).
\label{eq:cgf_V_ep2}
\end{align}

Following the same procedure as in Sec.~\ref{sec:LD}, we calculate $\mathcal{G}(h)$ to second order in $\epsilon$.
Using 
\begin{align}
\mathcal{L}_{h}^{(2)}&=\left( \mathcal{V}-\frac{\left( 8\eta^{2}+\pi\right)\mathcal{V}^{3}}{6\pi}\right)\frac{\partial}{\partial\mathcal{V}}-\left(2-\frac{3\mathcal{V}^{2}}{2}\right)\frac{\partial^{2}}{\partial\mathcal{V}^{2}}-\frac{8\mathcal{V}}{3}\frac{\partial^{3}}{\partial\mathcal{V}^{3}}+\frac{4\left( 8\eta^{2}+\pi\right)}{3\pi}\frac{\partial^{4}}{\partial\mathcal{V}^{4}},
\end{align}
the eigenvalue equation to second order in $\epsilon$ is written as
\begin{align}
\mathcal{L}_{h}^{(2)}\Psi_{h}^{(0)}+\mathcal{L}_{h}^{(1)}\Psi_{h}^{(1)}+\mathcal{L}_{h}^{(0)}\Psi_{h}^{(2)}=\mu_{h}^{(2)}\Psi_{h}^{(0)}+\mu_{h}^{(1)}\Psi_{h}^{(1)}+\mu_{h}^{(0)}\Psi_{h}^{(2)}.
\label{eq:ee_2}
\end{align}
The solvability condition for (\ref{eq:ee_2}) leads to
\begin{align}
\mu_{h}^{(2)}&=\frac{\left\langle \Phi_{h}^{(0)},\mathcal{L}_{h}^{(2)}\Psi_{h}^{(0)}\right\rangle}{\left\langle \Phi_{h}^{(0)},\Psi_{h}^{(0)}\right\rangle}+\frac{\left\langle \Phi_{h}^{(0)},\mathcal{L}_{h}^{(1)}\Psi_{h}^{(1)}\right\rangle}{\left\langle \Phi_{h}^{(0)},\Psi_{h}^{(0)}\right\rangle}-\mu_{h}^{(1)}\frac{\left\langle \Phi_{h}^{(0)},\Psi_{h}^{(1)}\right\rangle}{\left\langle \Phi_{h}^{(0)},\Psi_{h}^{(0)}\right\rangle}=\left(\frac{1}{2}+\frac{3\pi-8}{\pi}\eta^{2}\right) h^{2}+\left(\frac{2}{3}+8\eta^{2}\right)h^{4}.
\label{eq:mu_2}
\end{align}
Thus, we obtain
\begin{align}
\mathcal{G}(h) &= \epsilon\eta h+\left[ 1+\epsilon^{2}\left(\frac{1}{2}+\frac{3\pi-8}{\pi}\eta^{2}\right)\right] h^{2}+2\epsilon\eta h^{3}+\epsilon^{2}\left(\frac{2}{3}+8\eta^{2}\right)h^{4}+O(\epsilon^{3}).
\label{eq:Gh_ep2}
\end{align}

\section{Exact steady-state distribution for (\ref{eq:ELE_CGF})}\label{sec:app_NI}

The exact steady-state distribution of the Fokker--Planck equation corresponding to (\ref{eq:ELE_CGF}) is given by
\begin{align}
\hat{\mathcal{P}}_{\mathrm{ss}}^{\mathrm{ex}}(\hat{\mathcal{V}}) &=\mathcal{C}\left( 1-3\epsilon\eta\hat{\mathcal{V}}\right)^{\frac{1}{3\epsilon\eta}\left( \frac{1}{9\epsilon\eta}-2\epsilon\eta\right)} e^{-\frac{\hat{\mathcal{V}}^{2}}{3}+\frac{\hat{\mathcal{V}}}{9\epsilon\eta}},
\end{align}
where $\mathcal{C}$ is a normalization constant.
Because $\mathcal{C}$ is finite and $\hat{\mathcal{P}}_{\mathrm{ss}}^{\mathrm{ex}}(1/(3\epsilon\eta))=0$, the noise intensity of (\ref{eq:ELE_CGF}) is always positive during the time evolution.

\twocolumngrid

\end{document}